\documentclass[twocolumn,showpacs]{revtex4-1}
\usepackage{graphicx}
\usepackage{amsthm,amsmath,amssymb}
\usepackage{algorithm,algorithmic}
\usepackage{array}
\usepackage{color}
\usepackage{enumerate}
\usepackage{multirow}
\usepackage[normalem]{ulem}
\usepackage{subfig}

\newcommand{\nop}[1]{}
\newcommand{\Inorm}{I_\text{norm}}

\theoremstyle{definition}

\begin{document}
%\title{Z}
%\title{modularity with variance}
%\title{Normalizing modularity by the standard deviation.}
%\title{Z-score in community detection}
%\title{Finding community structure in networks using z-score}
%\title{Standardized modularity for community detection in networks}
%\title{Normalized modularity for community detection in networks}
\title{Z-score-based modularity for community detection in networks}
\author{Atsushi Miyauchi}
\email{miyauchi.a.aa@m.titech.ac.jp}
\author{Yasushi Kawase}
\email{kawase.y.ab@m.titech.ac.jp}
\affiliation{Graduate School of Decision Science and Technology, Tokyo Institute of Technology, Ookayama 2-12-1, Meguro-ku, Tokyo 152-8552, Japan}
\date{\today}
\begin{abstract}
Identifying community structure in networks is an issue of particular interest in network science. 
The modularity introduced by Newman and Girvan~[Phys. Rev. E \textbf{69}, 026113 (2004)] 
is the most popular quality function for community detection in networks.
%In this study, we discuss the normalization of the modularity using the standard deviation.
In this study, we identify a problem in the concept of modularity 
and suggest a solution to overcome this problem. 
Specifically, we obtain a new quality function for community detection. 
We refer to the function as \textit{Z-modularity} because
%our correction is based on the Z-score under some probability distribution.
it measures the Z-score of a given division with respect to the fraction of the number of edges within communities.
%As a result, it quantifies statistical rarity of community structure in terms of the fraction of the number of edges within communities. 
Our theoretical analysis shows that Z-modularity mitigates the resolution limit of the original modularity in certain cases.  
Computational experiments using both artificial networks and well-known real-world networks demonstrate 
the validity and reliability of the proposed quality function. 

\end{abstract}
%no more than 4 topics
%89.75.Hc: Genealogical trees (complex systems)
%02.70.-c: Computational techniques (mathematics)
%02.10.Ox: Combinatorics, Graph theory
%89.20.Hh: World Wide Web
%87.23.Ge: Social systems in ecology and evolution
%05.10.-a: Computational techniques (statistical physics and nonlinear dynamics)
\maketitle

\section{Introduction}\label{sec:introduction}
%ネットワーク科学の重要性
Many complex systems can be represented as networks. 
Analyzing the structure and dynamics of these networks provides meaningful information about the underlying systems. 
In fact, complex networks have attracted significant attention from diverse fields such as physics, informatics, chemistry, biology, and sociology~\cite{Ne03,Ne10}. 

%コミュニティの存在とその解析の重要性
An issue of particular interest in network science is the identification of community structure~\cite{Fo10}.  
Roughly speaking, a \textit{community} (also referred to as a \textit{module}) is a subset of vertices more densely connected with each other than with nodes in the rest of the network. 
Note that no absolute definition of a community exists because any such definition typically depends on the specific system at hand. 
Detecting communities is a powerful way to discover components that have some special roles or possess important functions. 
For example, consider the network representing the World Wide Web, where vertices correspond to web pages and edges represent the hyperlinks between pages. 
Communities in this network are likely to be the sets of web pages dealing with the same or similar topics. 

%コミュニティ検出法のクラス（特に評価関数によるアプローチ）
There are various methods to detect community structure in networks, which can be roughly divided into two types. First, there are methods based on some conditions that should be satisfied by a community. 
The most fundamental concept is a clique. 
A \textit{clique} is a subset of vertices wherein every pair of vertices is connected by an edge. 
As even a singleton or an edge is a clique, we are usually interested in finding a \textit{maximum clique} or a \textit{maximal clique}, 
i.e., cliques with maximum size and cliques not contained in any other clique, respectively. 
Although the definition of a clique is very intuitive, it is too strong and restrictive to use practically. 
In 2004, Radicchi \textit{et al.}~\cite{Raetal04} introduced more practical definitions: a community in a strong sense and a community in a weak sense. 
A subset $S$ of vertices is called a \textit{community in a strong sense} if for every vertex in $S$, the number of neighbors in $S$ is strictly greater than the number of neighbors outside $S$. 
On the other hand, a subset $S$ of vertices is called a \textit{community in a weak sense} if the sum, over all vertices in $S$, of the number of neighbors in $S$ is strictly greater than the number of cut edges of $S$. 
Thus, if a subset of vertices is a community in a strong sense, then it is also a community in a weak sense. 
Recently, Cafieri \textit{et al}.~\cite{Caetal12} proposed an enumerative algorithm to list all divisions of the set of vertices into communities in a strong sense with moderate sizes. 

Second, but perhaps more importantly, there are methods that maximize a globally defined quality function. 
The best known and most commonly used quality function is \textit{modularity}, 
which was introduced by Newman and Girvan~\cite{NeGi04}. 
Here let $G=(V, E)$ be an undirected network consisting of $n=|V|$ vertices and $m=|E|$ edges. 
The modularity, a quality function for division $\mathcal{C}=\{C_1,\dots,C_k\}$ of $V$ 
(i.e., $\bigcup_{i=1}^k C_i = V$ and $C_i\cap C_j = \emptyset$ for $i\neq j$), can be written as 
\begin{align*}
Q(\mathcal{C})=\sum_{C\in\mathcal{C}}\left(\frac{m_C}{m}-\left(\frac{D_C}{2m}\right)^2\right),
\end{align*}
where $m_C$ is the number of edges in community $C$, and $D_C$ is the sum of the degrees of the vertices in community $C$. 
The modularity represents the sum, over all communities, of the fraction of the number of edges in the communities 
minus the expected fraction of such edges assuming that they are placed at random with the same distribution of vertex degree. 

%モジュラリティ最大化の既存研究（計算複雑度，種々の解法）
Many studies have examined modularity maximization. 
In 2008, Brandes \textit{et al.}~\cite{Bretal08} proved that modularity maximization is NP-hard. 
This implies that unless P = NP, no modularity maximization method that simultaneously satisfies the following exists: 
(i) finds a division that maximizes modularity exactly (ii) in time polynomial in $n$ and $m$ (iii) for any networks. 
To date, a major focus in modularity maximization has been designing accurate and scalable heuristics. 
In fact, there are a wide variety of algorithms based on greedy techniques~\cite{Bletal08,ClNeMo04,NeGi04}, simulated annealing~\cite{GuAm05,MaDo05,MeAcDo05}, extremal optimization~\cite{DuAr05}, spectral optimization~\cite{Ne06-1,RiMuPo09}, mathematical programming~\cite{AgKe08,CaHaLi11,MiMi13,CaCoHa14}, and other techniques. 
Note that to reduce computation time, a few pre-processing techniques have been proposed~\cite{Aretal07}. Moreover, to improve the quality of divisions obtained by such heuristics, some post-processing algorithms have also been developed~\cite{CaHaLi14}. 

%モジュラリティの問題点（resolution limit and degeneracies）
Although modularity maximization is the most popular and widely used method in practice, it is also known to have 
some serious drawbacks; i.e., the \textit{resolution limit}~\cite{FoBa07} and \textit{degeneracies}~\cite{GoMoCl10}. 
The former means that modularity maximization fails to detect communities smaller than a certain scale 
depending on the total number of edges in a network even if the communities are cliques connected by single edges. 
The latter means that there exist numerous nearly optimal divisions in terms of modularity maximization, 
which makes finding communities with maximum modularity extremely difficult. 
The resolution limit particularly narrows the application range of modularity maximization 
because most real-world networks consist of communities with very different sizes. 
To avoid this issue, some multiresolution variants of the modularity have been adopted 
in practical applications~\cite{ReBo06,ArFeGo08,PoLa11}. 
In these variants, the resolution level can be tuned freely by adjusting certain parameters.  
However, once the resolution level is determined, communities larger than the determined resolution level tend to be divided 
and smaller communities tend to be merged. 
Therefore, such multiresolution variants also fail to detect real community structure~\cite{LaFo11}.

%他の評価関数によるコミュニティ検出
%modularity density~\cite{Lietal08}
%Max-Min modularity
%surprise~\cite{AlMa10,AlMa11}

%本研究の概要（大雑把に）

%In this study, we propose a new quality function for community detection in networks. 
%Conceptually speaking, our function quantifies statistical rarity of community structure 
%by standardizing the modularity using the standard deviation. 
%It should be emphasized that our function mitigates the resolution limit of the modularity in certain cases.  
%Indeed, it correctly identifies natural community structure of the well-known \textit{ring of cliques network} 
%with any number and size of cliques. 
%Our computational experiments using both artificial networks and famous real-world networks show 
%the validity and reliability of our quality function. 

%% In this study, we discuss the normalization of the modularity using the standard deviation. 
%% As a result, we obtain a new quality function for community detection in networks. 
%% We call the quality function \textit{Z-modularity} since it measures Z-score of a given division with respect to the fraction of the number of edges within communities. 
%Conceptually speaking, Z-modularity quantifies statistical rarity of community structure.
In this study, we identify a problem in the concept of modularity and suggest a solution to overcome this problem.
Specifically, we obtain a new quality function for community detection. 
We refer to this function as \textit{Z-modularity} because
it measures the Z-score of a given division with respect to the fraction of the number of edges within communities.
Our theoretical analysis shows that Z-modularity mitigates the resolution limit of the original modularity in certain cases. 
%It should be emphasized that our quality function mitigates the resolution limit of the original modularity in certain cases.  
%Indeed, it correctly identifies natural community structure of the well-known ring of cliques network 
In fact, Z-modularity never merges  adjacent cliques in the well-known ring of cliques network with any number and size of cliques. %%%%%%%added
Computational experiments using both artificial networks and well-known real-world networks demonstrate
the validity and reliability of the proposed quality function.

Note that there are many quality functions based on modularity 
or other concepts~\cite{RoBe07,Lietal08,RoBe08,ChZaGo09,ZhZh12,ZhZh13}.
%For more details, see Ref.~\cite{Fo10}.
Most of them are collected in Ref.~\cite{Fo10}.
%To the best of our knowledge, there exists no quality function that is based on the Z-score. 
%% In particular, Zhang and Zhao~\cite{ZhZh13} introduced the \textit{normalized modularity with degree adjustment}, 
%% which normalizes the original modularity by taking into account the average degree of communities instead of the number of edges within communities. 
%% For more details, see Ref.~\cite{Fo10}. 
%% To the best of our knowledge, there exists no quality function that normalizes the modularity using statistical techniques. 

%We note that there have been already some measures 
%to quantify the statistical significance of community structure~\cite{LaRaRa10,Raetal10,Laetal11}. 
%However, they depend on the hypergeometric distribution and thus the computational burden is too heavy. 
%In contrast to them, our quality function can be computed 
%as easily as the original modularity due to some reasonable approximation used in the definition. 
%For more details, see Ref.~\cite{Fo10}. 
%However, to the best of our knowledge, no quality function based on the normalization of the modularity exists. 

%\TODO{add more detailed explanation}

%論文の構成
This paper is structured as follows. 
In Sec.~\ref{sec:model}, our quality function Z-modularity is introduced. 
In Sec.~\ref{sec:analysis}, a theoretical analysis of the properties of Z-modularity is described. 
The results of computational experiments are shown in Sec.~\ref{sec:experiments}. 
Finally, conclusions and suggestions for future work are given in Sec.~\ref{sec:conclusion}. 

\section{Definition of Z-modularity}\label{sec:model}

Modularity simply computes the fraction of the number of edges within communities minus its expected value. 
The definition is quite intuitive; thus, it is the most popular and widely used quality function in practice. 

\begin{figure}[tb]
\centering
\includegraphics[width=0.4 \textwidth]{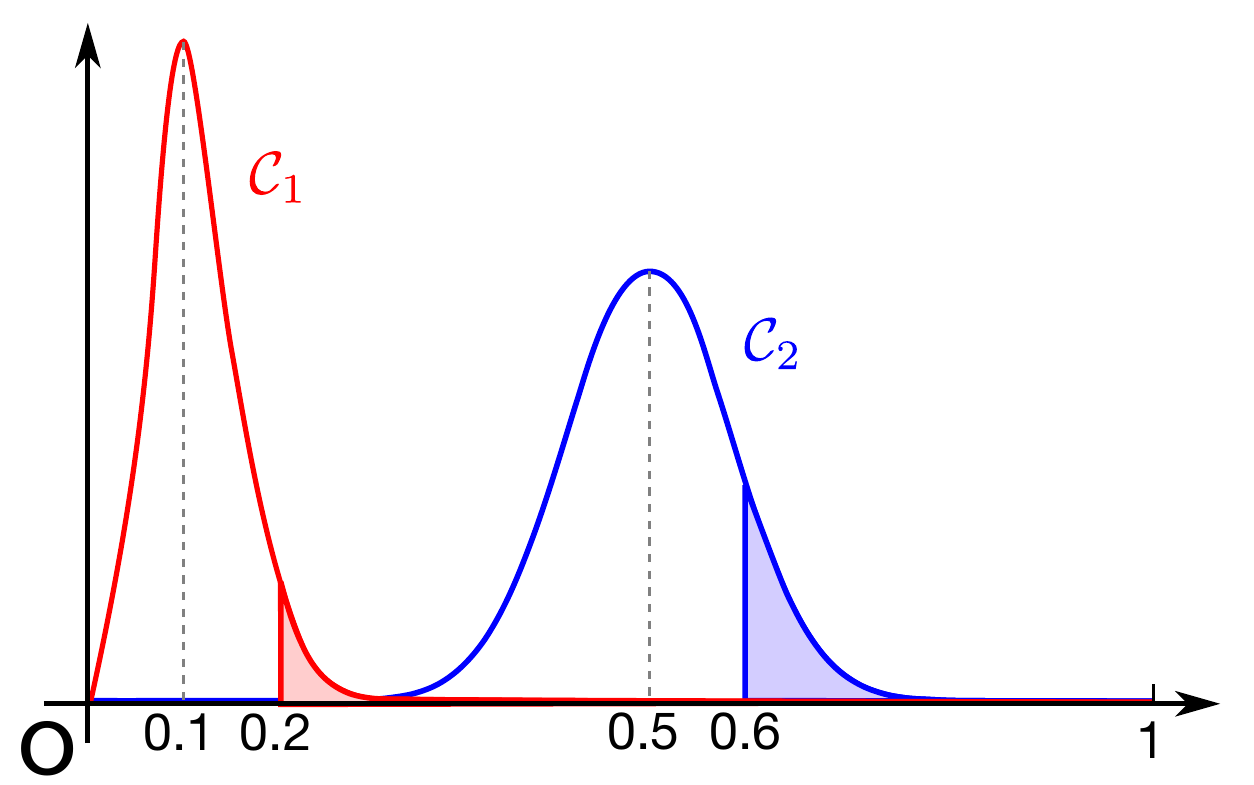}
\caption{(Color online) Probability distributions.} \label{fig:prob-dist}
\end{figure}

However, we identify a problem with the concept of modularity.  
Here consider two divisions $\mathcal{C}_1$ and $\mathcal{C}_2$. 
Assume that the fraction of the number of edges within communities of $\mathcal{C}_1$ and $\mathcal{C}_2$ 
are 0.2 and 0.6, respectively. 
In addition, assume that their expected values are 0.1 and 0.5, respectively. 
Then, we see that these two divisions share the same modularity value (i.e., $Q(\mathcal{C}_1) = Q(\mathcal{C}_2) = 0.1$). 
The key question is as follows: 
should these two divisions receive the same quality value? 
Our answer is that it must depend on the variance of the probability distribution of the fraction of the number of edges 
within communities of $\mathcal{C}_1$ and $\mathcal{C}_2$. 
Fig.~\ref{fig:prob-dist} illustrates an example. In this case, we wish to assign a higher quality value to $\mathcal{C}_1$ 
because it is statistically much rarer than $\mathcal{C}_2$. 
This simple but critical observation forms the basis of our quality function. 

%We now pose a question. Which division is statistically rarer? 
%The situation is described in Fig.~\ref{fig:prob-dist}. 
%This simple but critical observation forms the basis of our quality function. 

Given an undirected network $G=(V,E)$ consisting of $n=|V|$ vertices and $m=|E|$ edges, and a division $\mathcal{C}$ of $V$, 
we aim to quantify the statistical rarity of division $\mathcal{C}$ in terms of the fraction of the number of edges within communities. 
To this end, we consider the following edge generation process over $V$. 
Place $N$ edges over $V$ at random with the same distribution of vertex degree. 
%When we put an edge, the probability of having $i, j \in V$ connected is $d_id_j/(2m)^2$.
Then, when we place an edge, the probability that the edge is placed within communities is given by 
\begin{align*}
p = \sum_{C\in \mathcal{C}} \left(\frac{D_C}{2m} \right)^2. 
\end{align*}
Note that this edge generation process is the same as the null-model (also known as the \textit{configuration model}~\cite{MoRe95}) 
used in the definition of modularity, with the exception of the sample size. 
We simply wish to estimate the probability distribution of the fraction of the number of edges within communities. 
Thus, unlike the null-model, the sample size $N$ is not necessarily equal to the number of edges $m$. 

Let $X$ be a random variable denoting the number of edges generated by the process within communities. 
Then, $X$ follows the binomial distribution $B(N, p)$. 
%de Moivre-Laplace theorem???
By the central limit theorem, when the sample size $N$ is sufficiently large, 
the distribution of $X/N$ can be approximated by the normal distribution $\mathcal{N}(p,p(1-p)/N)$. 
Thus, we can quantify the statistical rarity of division $\mathcal{C}$ 
in terms of the fraction of the number of edges within communities using the Z-score as follows: 
\begin{align*}
%Z(\mathcal{C})=\frac{\sum_{C\in\mathcal{C}}\left(\frac{m_C}{m}-\left(\frac{D_C}{2m}\right)^2\right)}{\sqrt{\sum_{C\in\mathcal{C}}\left(\frac{D_C}{2m}\right)^2 \left(1-\sum_{C\in\mathcal{C}}\left(\frac{D_C}{2m}\right)^2\right)}}.
Z(\mathcal{C})=\frac{\sum_{C\in\mathcal{C}}\frac{m_C}{m}-\sum_{C\in\mathcal{C}}\left(\frac{D_C}{2m}\right)^2}{\sqrt{\sum_{C\in\mathcal{C}}\left(\frac{D_C}{2m}\right)^2 \left(1-\sum_{C\in\mathcal{C}}\left(\frac{D_C}{2m}\right)^2\right)}}.
\end{align*}
The sample size $N$ never depends on a given division; thus, it is omitted in the denominator. We refer to this quality function as \textit{Z-modularity}.

\if 0
Let $X$ be a random variable denoting the number of edges generated by the process within communities. 
Then, $X$ follows the binomial distribution $B(N, p)$. 
%de Moivre-Laplace theorem???
By the central limit theorem, when the sample size $N$ is sufficiently large, 
the distribution of $X/N$ can be approximated by the normal distribution $\mathcal{N}(p,p(1-p)/N)$. 
Thus, we can quantify the statistical significance of division $\mathcal{C}$ in terms of the fraction of the number of edges within communities using \textit{z-score} as follows: 
\begin{align*}
%\Zm(\mathcal{C})=\frac{\sum_{C\in\mathcal{C}}\left(\frac{m_C}{m}-\left(\frac{D_C}{2m}\right)^2\right)}{\sqrt{\left(\sum_{C\in\mathcal{C}}\left(\frac{D_C}{2m}\right)^2\right)\left(1-\sum_{C\in\mathcal{C}}\left(\frac{D_C}{2m}\right)^2\right)}}.
  \frac{\sum_{C\in\mathcal{C}}\frac{m_C}{m} - p}{\sqrt{p(1-p)/N}}.
  %\frac{\sum_{C\in\mathcal{C}}m_C - m\cdot p}{m\sqrt{p(1-p)/N}}=\frac{Q(\mathcal{C})}{\sqrt{p(1-p)/N}}.
\end{align*}
%For simplicity, we omit $N$ and call the following quality function \textit{Z-modularity}:
Since the sample size $N$ does not depend on a given division, we can omit it in the denominator as follows: 
\begin{align*}
  %Z(\mathcal{C})=\frac{Q(\mathcal{C})}{\sqrt{p(1-p)}}.
  Z(\mathcal{C})=\frac{\sum_{C\in\mathcal{C}}\frac{m_C}{m} - p}{\sqrt{p(1-p)}}.
\end{align*}
We call this quality function \textit{Z-modularity}. 
\fi
\section{Theoretical analysis}\label{sec:analysis}
%\section{Resolution limit of Z-modularity}\label{sec:analysis}
Fortunato and Barth{\'e}lemy~\cite{FoBa07} pointed out the resolution limit of modularity. 
This resolution limit means that modularity maximization fails to detect communities that are smaller than a certain scale depending on the total number of edges in a network. 
This phenomenon occurs even if the communities are cliques connected by single edges. 
Here we theoretically analyze Z-modularity from a resolution limit perspective. 
%Specifically, we discuss the resolution limit of Z-modularity. 
%Specifically, we discuss Z-modularity from the resolution limit point of view. 
As a result, we demonstrate that Z-modularity mitigates the resolution limit of the original modularity in certain cases. 
%Fortunato and Barth{\'e}lemy~\cite{FoBa07}
%modularity maximization fails to detect communities smaller than a certain scale depending on the total number of edges in a network. 

\subsection{Ring of cliques network}
\begin{figure}[tb]
\centering
\includegraphics[width=0.29 \textwidth]{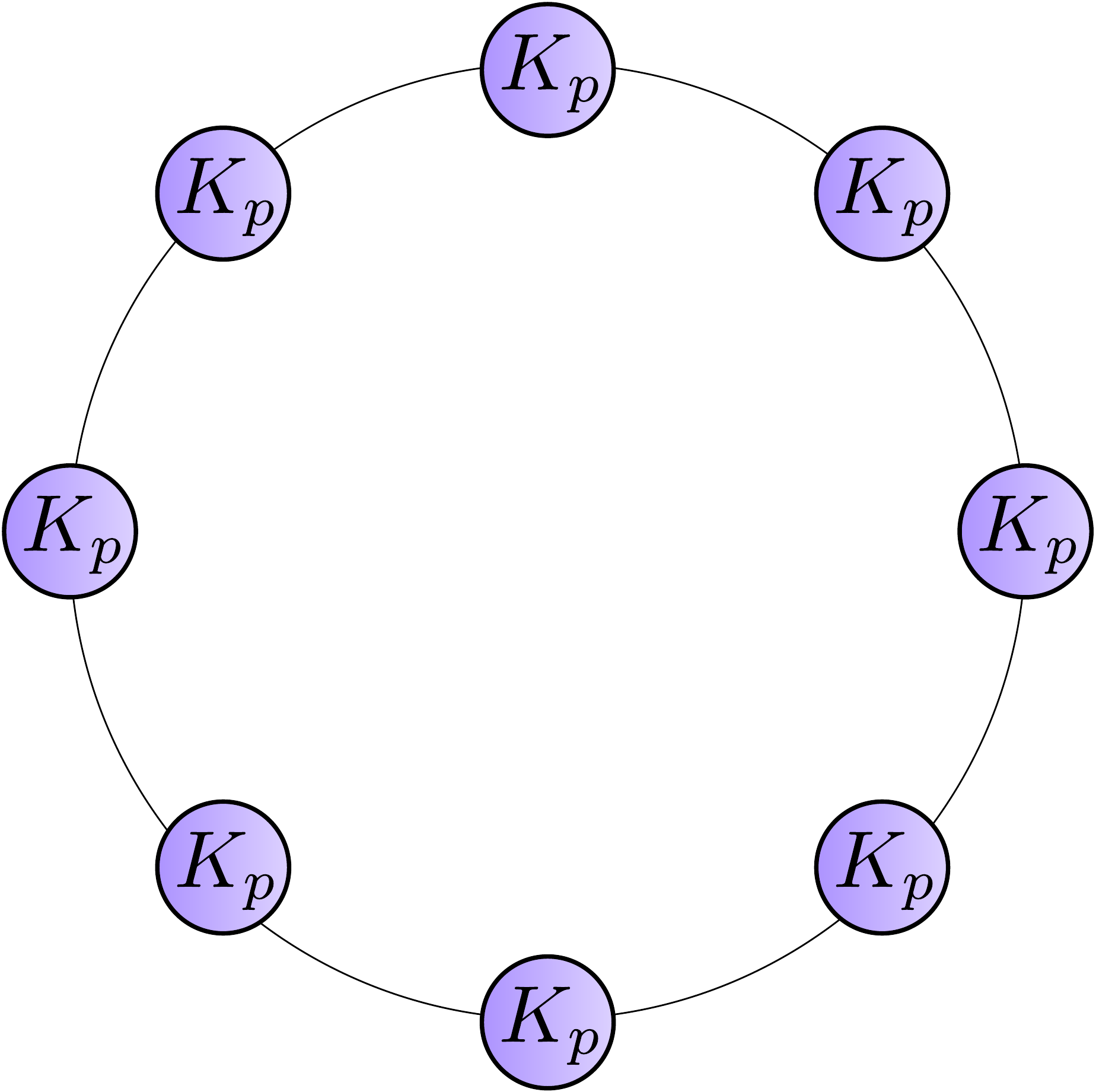}
\caption{(Color online) Ring of cliques network. $K_p$ represents a clique with $p$ vertices.} \label{fig:cliquering}
\end{figure}

First, we consider a \textit{ring of cliques network} that consists of a number of cliques connected by single edges (Fig.~\ref{fig:cliquering}).
Assume that each clique consists of $p\ (\geq 3)$ vertices and the number of cliques is $q\ (\geq 2)$.
Then, the network has $n=p\cdot q$ vertices and $m=q\cdot(1+p(p-1)/2)$ edges.
Fortunato and Barth{\'e}lemy~\cite{FoBa07} showed that
modularity maximization would merge adjacent cliques if $q$ is larger than a certain value depending on $p$.
%However, in a division with maximal Z-modularity value, adjacent cliques are never merged, as shown below.
However, adjacent cliques are never merged in a division with maximal Z-modularity value, as shown below.

Let $\mathcal{C}^*$ be the division of $V$ into the cliques. 
In addition, let $\mathcal{C}=\{C_1,\dots,C_l\}$ ($1<l<q$) be a division of $V$ 
such that each $C_i$ consists of a series of $s_i\ (\geq 1)$ cliques and $q=\sum_{i=1}^l s_i$. 
Then, Z-modularity for $\mathcal{C}^*$ and $\mathcal{C}$ are calculated by
\begin{align*}
Z(\mathcal{C}^*)=\frac{1-q/m-1/q}{\sqrt{(1-1/q)/q}}\quad\text{and}\quad
Z(\mathcal{C})=\frac{1-l/m-t}{\sqrt{t(1-t)}},
\end{align*}
%\begin{align*}
%Z(\mathcal{C}^*)=\frac{1-q/m-1/q}{\sqrt{(1-1/q)/q}}
%\end{align*}
%and
%\begin{align*}
%Z(\mathcal{C})=\frac{1-l/m-t}{\sqrt{t(1-t)}},
%\end{align*}
respectively, where $t=\sum_{i=1}^l(s_i/q)^2$.
By the Cauchy--Schwarz inequality, we have $1>t=\sum_{i=1}^l(s_i/q)^2\ge \left(\sum_{i=1}^l (s_i/q)\right)^2/l=1/l$. 
Here define
\begin{align*}
f(x,y)=\frac{1-y/m-x}{\sqrt{x(1-x)}}.
\end{align*}
Then, the derivative of $f(x,y)$ with respect to $x$ is 
\begin{align*}
  \frac{\partial}{\partial x}f(x,y)&=\frac{-x\cdot y/m-(1-y/m)(1-x)}{2\cdot (x(1-x))^{3/2}}<0
\end{align*}
for $0<x<1$ and $1\le y\le m$. Thus, we obtain 
\begin{align*}
f(1/l,l)\ge f(t,l).
\end{align*}
%Moreover, the derivative of $f(1/l,l)$ with respect to $l$ is 
Moreover, the derivative of $f(1/y,y)$ with respect to $y$ is 
\begin{align*}
  \frac{\partial}{\partial y}f(1/y,y)
  &=\frac{(m-3y)(y-1)+y}{2m\cdot (y-1)^{3/2}}>0
\end{align*}
for $1< y<m/3$.
Thus, we have 
\begin{align*}
f(1/q,q)>f(1/l,l),
\end{align*}
since $1<l< q\le m/4$ by $m=q\cdot (1+p(p-1)/2)\ge 4q$.
Therefore, we have
\begin{align*}
Z(\mathcal{C}^*)=f(1/q,q)>f(1/l,l)\ge f(t,l)=Z(\mathcal{C}),
\end{align*}
%for any division $\mathcal{C}$, 
which means that maximizing Z-modularity never merges adjacent cliques. 

Table~\ref{tab:Kring} lists the values of modularity and Z-modularity 
of  divisions $\mathcal{C}^*$ and $\mathcal{C}$ ($s_i=2$ for $i=1,\dots,l$) for some ring of cliques networks. 
As can be seen, 
the modularity of $\mathcal{C}$ is greater than that of $\mathcal{C}^*$ when the number of cliques is large, 
which is consistent with Fortunato and Barth{\'e}lemy~\cite{FoBa07}. 
On the other hand, as we proved above, Z-modularity of $\mathcal{C}^*$ is certainly higher than that of $\mathcal{C}$ 
for every number of cliques.

%Table~\ref{tab:Kring} lists the values of modularity and Z-modularity 
%of  divisions $\mathcal{C}^*$ and $\mathcal{C}_{q/2}$ for some ring of cliques networks. 
%As can be seen, 
%the modularity of $\mathcal{C}_{q/2}$ is greater than that of $\mathcal{C}^*$ when the number of cliques is large, 
%which is consistent with Fortunato and Barth{\'e}lemy~\cite{FoBa07}. 
%On the other hand, as we proved above, Z-modularity of $\mathcal{C}^*$ is certainly higher than that of $\mathcal{C}_{q/2}$ 
%for every number of cliques.

%We also compare the values of Z-modularity and the modularity
%with respect to the divisions $\mathcal{C}_q$ ($k=1$) and $\mathcal{C}_{q/2}$ ($k=2$) 
%for some ring of cliques networks.
%The results are listed in Table~\ref{tab:Kring}. 
%They show that Z-modularity of the case $k=1$ is certainly higher than
%that of $k=2$ for every number of cliques $q$.

% n=p*k*l
% m=k*l*(1+p(p-1)/2)
\begin{table}[tb]
%\caption{\label{tab:Kring}Numerical examples of Z-modularity and modularity for ring of cliques networks.}
\caption{\label{tab:Kring}Numerical examples of modularity and Z-modularity for some ring of cliques networks.}
\begin{ruledtabular}
\begin{tabular}{rrrrrrrr}
$n$ & $m$ & $p$ & $q$ & $Q(\mathcal{C}^*)$ & $Q(\mathcal{C})$& $Z(\mathcal{C}^*)$ & $Z(\mathcal{C})$\\\hline
100 & 220 & 5& 20& \textbf{0.8591} & 0.8548& \textbf{3.942} & 2.848 \\
200 & 440 & 5& 40& 0.8841 & \textbf{0.9045}& \textbf{5.663} & 4.150 \\
400 & 880 & 5& 80& 0.8966 & \textbf{0.9295}& \textbf{8.070} & 5.954 \\
5000&11000& 5&1000& 0.9081 & \textbf{0.9525}&\textbf{28.73} & 21.32 \\
\end{tabular}
%\begin{tabular}{rrrrrrrr}
%$n$ & $m$ & $p$ & $q$ & $Z(\mathcal{C}_q)$ & $Z(\mathcal{C}_{q/2})$ & $Q(\mathcal{C}_q)$ & $Q(\mathcal{C}_{q/2})$\\\hline
%100 & 220 & 5& 20& \textbf{3.942} & 2.848 & \textbf{0.8591} & 0.8548\\
%200 & 440 & 5& 40& \textbf{5.663} & 4.150 & 0.8841 & \textbf{0.9045}\\
%400 & 880 & 5& 80& \textbf{8.070} & 5.954 & 0.8966 & \textbf{0.9295}\\
%5000&11000& 5&1000&\textbf{28.73} & 21.32 & 0.9081 & \textbf{0.9525}\\
%\end{tabular}
%% \begin{tabular}{rrrrrrrr}
%% $n$ & $m$ & $p$ & $q$ & $l$ & $k$ & $Z$  & $Q$\\\hline
%% %% 5& 10& 1   &  \textbf{2.70} & \textbf{0.809} \\ 
%% %% 5& 10& 2   &  1.89          & 0.754 \\
%% 100 & 220 & 5& 20& 20 & 1   &  \textbf{3.94} & \textbf{0.859} \\
%% 100 & 220 & 5& 20& 10 & 2   &  2.85 & 0.855 \\
%% \noalign{\vskip 1mm}
%% 200 & 440 & 5& 40& 40 & 1   &  \textbf{5.66} & 0.884 \\
%% 200 & 440 & 5& 40& 20 & 2   &  4.15 & \textbf{0.905} \\
%% \noalign{\vskip 1mm}
%% 400 & 880 & 5& 80& 80 & 1   &  \textbf{8.07} & 0.897 \\
%% 400 & 880 & 5& 80& 40 & 2   &  5.95 & \textbf{0.930} \\
%% \noalign{\vskip 1mm}
%% 5000& 11000 & 5& 1000& 5000 & 1 &  \textbf{28.7} & 0.908 \\
%% 5000& 11000 & 5& 1000& 2500 & 2   &  21.3 & \textbf{0.953} \\
%% %% 100& 100& 1   &  \textbf{9.95} & \textbf{0.990} \\
%% %% 100& 100& 2   &  7.00 & 0.980 \\
%% %% 100& 10000& 1   &  \textbf{100} & 0.999698 \\
%% %% 100& 10000& 2   &  70.6 & \textbf{0.999699} \\
%% \end{tabular}
\end{ruledtabular}
\end{table}

\subsection{Network with two pairwise identical cliques}
\begin{figure}[tb]
\centering
\includegraphics[width=0.28 \textwidth]{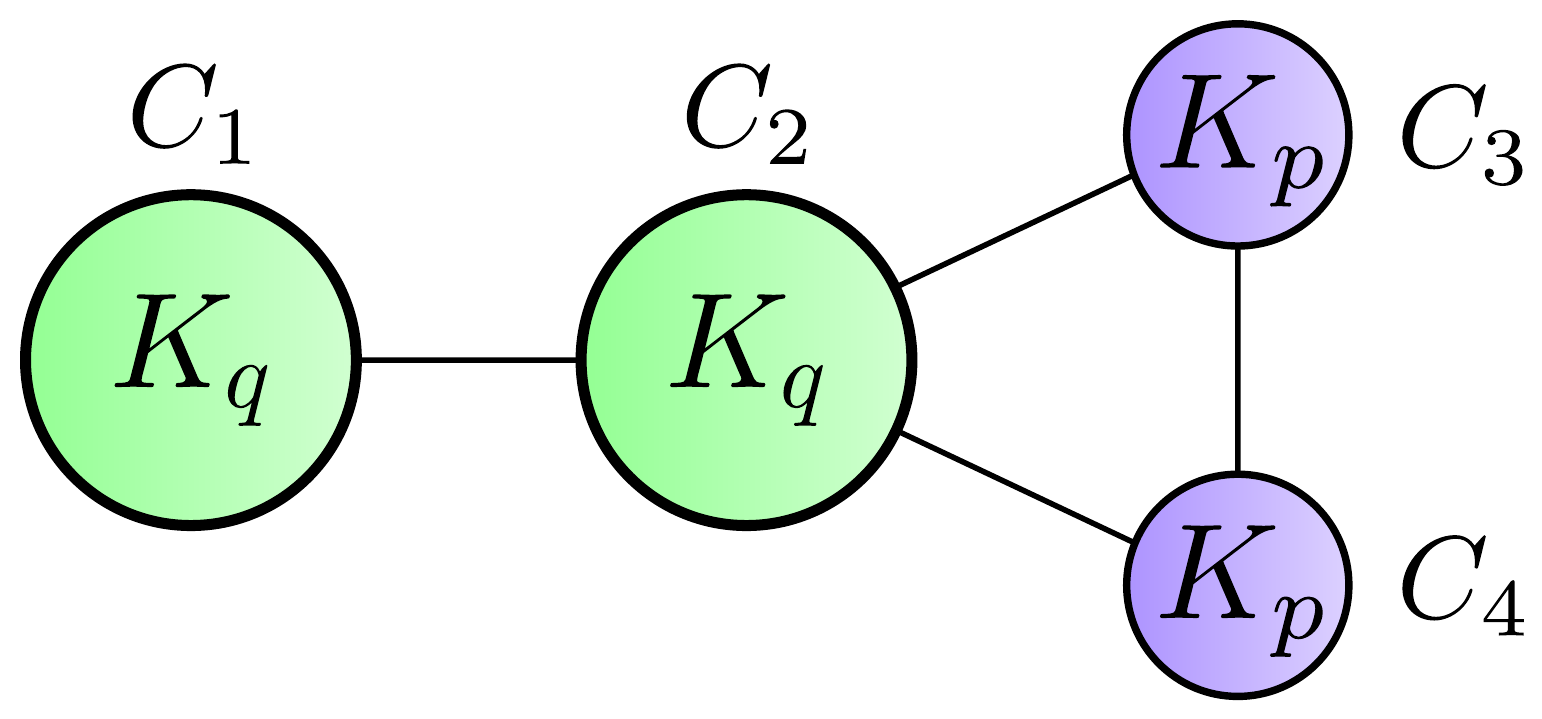}
\caption{(Color online) Network with two pairwise identical cliques. $K_p$ and $K_q$ represent cliques with $p$ and $q$ vertices, respectively.} \label{fig:2pcliques}
\end{figure}

Here we consider a \textit{network with two pairwise identical cliques} that consists of a pair of cliques $C_1$ and $C_2$ with $q$ vertices each and a pair of cliques $C_3$ and $C_4$ with $p\ (< q)$ vertices each.
These four cliques are connected by single edges, as described in Fig.~\ref{fig:2pcliques}.
This network has $n=2(p+q)$ vertices and $m=p(p-1)+q(q-1)+4$ edges.

Consider two divisions $\mathcal{C}_A=\{C_1,C_2,C_3,C_4\}$ and $\mathcal{C}_B=\{C_1,C_2,C_3\cup C_4\}$.
%Needless to say,
Note that division $\mathcal{C}_A$ is more natural community structure that we would like to identify. 
%\TODO{explain why $\mathcal{C}_A$ should be chosen}
Unfortunately, maximizing Z-modularity may choose $\mathcal{C}_B$, 
i.e., $Z(\mathcal{C}_A)<Z(\mathcal{C}_B)$ holds for some pair of $p$ and $q$.
However, if modularity maximization adopts $\mathcal{C}_A$, then so does Z-modularity, 
i.e., for any pair of $p$ and $q$, if $Q(\mathcal{C}_A)>Q(\mathcal{C}_B)$ holds, then $Z(\mathcal{C}_A)>Z(\mathcal{C}_B)$ also holds. 
This fact follows from the definitions of Z-modularity and the original modularity. 

Table~\ref{tab:2pcliques} lists the values of modularity and Z-modularity of divisions $\mathcal{C}_A$ and $\mathcal{C}_B$ 
for some networks with two pairwise identical cliques. 
We can confirm that 
%due to the resolution limit, 
both modularity and Z-modularity tend to merge $C_3$ and $C_4$ as the sizes of $C_1$ and $C_2$ become large.
However, there is the case where only Z-modularity could divide $C_3$ and $C_4$. 
Therefore, we see that Z-modularity again mitigates the resolution limit of modularity in this case. 

%We also compare the values of Z-modularity and modularity for some networks with two pairwise identical cliques when $p=5$.
%The results are listed in Table~\ref{tab:2pcliques}, which show that the properties discussed above.

\begin{table}[tb]
%\caption{\label{tab:2pcliques}Numerical examples of Z-modularity and modularity for the networks with two pairwise identical cliques.}
\caption{\label{tab:2pcliques}Numerical examples of modularity and Z-modularity for some networks with two pairwise identical cliques.}
\begin{ruledtabular}
\begin{tabular}{rrrrrrrr}
  $n$ & $m$ & $p$ & $q$  & $Q(\mathcal{C}_A)$ & $Q(\mathcal{C}_B)$& $Z(\mathcal{C}_A)$  & $Z(\mathcal{C}_B)$\\\hline
  26  & 80  &   5 &  8 &    \textbf{0.6618}  & 0.3385       & \textbf{1.443}        & 1.345               \\
  42  & 264 &   5 & 16        & 0.5650           & \textbf{0.5653}     & \textbf{1.144}     & 1.143             \\
  74  &1016 &   5 & 32   & 0.5182           & \textbf{0.5190}      & 1.037              & \textbf{1.039}        \\
  138 &4056 &   5 & 64    & 0.5047           & \textbf{0.5049}      & 1.009              & \textbf{1.010}       \\
\end{tabular}
%\begin{tabular}{rrrrrrrr}
%  $n$ & $m$ & $p$ & $q$ & $Z(\mathcal{C}_A)$  & $Z(\mathcal{C}_B)$ & $Q(\mathcal{C}_A)$ & $Q(\mathcal{C}_B)$\\\hline
%  26  & 80  &   5 &  8 & \textbf{1.4432}     & 1.3451            & \textbf{0.66180}  & 0.33852             \\
%  42  & 264 &   5 & 16 & \textbf{1.1435}     & 1.1429            & 0.56502           & \textbf{0.56533}             \\
%  74  &1016 &   5 & 32 & 1.0374              & \textbf{1.0389}   & 0.51823           & \textbf{0.51898}             \\
%  138 &4056 &   5 & 64 & 1.0094              & \textbf{1.0098}   & 0.50466           & \textbf{0.50489}             \\
%\end{tabular}
\end{ruledtabular}
\end{table}

\if 0
\begin{table}[tb]
\caption{\label{tab:2pcliques}Numerical examples of Z-modularity and modularity for the networks with two pairwise identical cliques.}
\begin{ruledtabular}
\begin{tabular}{rrrrrrrr}
  $n$ & $m$ & $p$ & $q$ & $Z(\mathcal{C}_A)$  & $Z(\mathcal{C}_B)$ & $Q(\mathcal{C}_A)$ & $Q(\mathcal{C}_B)$\\\hline
  26  & 80  &   5 &  8 & \textbf{1.44319}     & 1.34505            & \textbf{0.661797}  & 0.338516             \\
  42  & 264 &   5 & 16 & \textbf{1.14346}     & 1.14288            & 0.565018           & \textbf{0.565334}             \\
  74  &1016 &   5 & 32 & 1.03739              & \textbf{1.03887}   & 0.518231           & \textbf{0.518981}             \\
  138 &4056 &   5 & 64 & 1.00937              & \textbf{1.00983}   & 0.504655           & \textbf{0.504887}             \\
\end{tabular}
\end{ruledtabular}
\end{table}
\fi

\section{Experimental Results}\label{sec:experiments}
%The purpose of our computational experiments is to evaluate the validity and reliability of the quality function Z-modularity. Thus, if possible, some exact algorithm should be adopted. However, how to design an exact algorithm other than the brute-force search is not obvious. Hence, we optimize Z-modularity on artificial networks and famous real-world networks. Throughout the experiments, we use the simulated annealing algorithm

The purpose of our computational experiments is to evaluate the validity and reliability of the quality function Z-modularity. 
To this end, throughout the experiments, we maximize Z-modularity using a simulated annealing algorithm. 
Note that our algorithm is obtained immediately by changing the objective function from modularity to Z-modularity 
in the algorithm proposed by Guimer{\`a} and Amaral~\cite{GuAm05}. 
The implementation of their algorithm can be found on Lancichinetti's web page~\cite{La}, 
and we use it with default parameters with the exception of the above change of objective function. 
Our experiments are conducted on various artificial networks and on well-known real-world networks. 

\subsection{Artificial networks}
First, we report the results of computational experiments with artificial networks. 
We compare divisions obtained by maximizing Z-modularity with divisions obtained by modularity maximization 
on a wide variety of networks. 
The modularity is also maximized by the simulated annealing algorithm proposed by Guimer{\`a} and Amaral~\cite{GuAm05}. 
We deal with three types of artificial networks: the planted $l$-partition model, the Lancichinetti--Fortunato--Radicchi (LFR) benchmark, and the Hanoi graph. 
For the planted $l$-partition model and the LFR benchmark, once their parameters are set, 
the ground-truth community structure is fixed. 
Thus, we can evaluate the quality of the obtained community structure by comparison with the ground-truth using some measure. 
%Hence, we can evaluate the quality of obtained community structure by computing the similarity between the division and the ground-truth in terms of some measure. 

To this end, we adopt the \textit{normalized mutual information} introduced by Danon \textit{et al.}~\cite{Daetal05}. 
The normalized mutual information for two divisions $\mathcal{C}_1$ and $\mathcal{C}_2$
of $n$ vertices is defined as follows:
\begin{align*}
\Inorm(\mathcal{C}_1,\mathcal{C}_2)=\frac{2I(\mathcal{C}_1,\mathcal{C}_2)}{H(\mathcal{C}_1)+H(\mathcal{C}_2)},
\end{align*}
where 
\begin{align*}
I(\mathcal{C}_1,\mathcal{C}_2)=\sum_{C_1\in\mathcal{C}_1}\sum_{C_2\in\mathcal{C}_2}
\frac{|C_1\cap C_2|}{n}\log_2\left(\frac{n\cdot |C_1\cap C_2|}{|C_1|\cdot |C_2|}\right)
\end{align*}
and 
\begin{align*}
H(\mathcal{C})=-\sum_{C\in\mathcal{C}}\frac{|C|}{n}\log_2\frac{|C|}{n}.
\end{align*}
The normalized mutual information ranges from 0 to 1. 
For two divisions $\mathcal{C}_1$ and $\mathcal{C}_2$, 
the higher the normalized mutual information is, the more similar they are (and vice versa). 
In fact, $\Inorm(\mathcal{C}_1,\mathcal{C}_2)=1$ if $\mathcal{C}_1$ and $\mathcal{C}_2$ are identical,
and $\Inorm(\mathcal{C}_1,\mathcal{C}_2)=0$ if they are independent.
This measure has often been used to evaluate community detection methods. 
For example, see the computational experiments in Refs.~\cite{LaFo09,LaFo14}.

\medskip
\paragraph*{Planted $l$-partition model.}

The planted $l$-partition model was introduced by Condon and Karp~\cite{CoKa01}. 
In this model, $n$ vertices are divided into $l$ equally sized groups. 
Two vertices in the same group are connected by probability $p_\text{in}$, whereas
two vertices in different groups are connected by probability $p_\text{out}~(<p_\text{in})$.
Throughout the experiments, we set $p_\text{in}=0.5$. 
We construct four networks corresponding to combinations of two different network sizes ($n=1000$ or 5000) 
and two different community sizes ($l=20$ or 50). 
%In our experiments, we set the parameters as follows:
%(i) $n=1000$, $l=20$, and $p_\text{in}=0.5$;
%(ii) $n=1000$, $l=50$, and $p_\text{in}=0.5$;
%(iii) $n=5000$, $l=20$, and $p_\text{in}=0.5$;
%(iv) $n=5000$, $l=50$, and $p_\text{in}=0.5$.
The parameter $p_\text{out}$ starts with 0.01 and then increases in stages. 
%For each $p_\text{out}$, 
%we compare the normalized mutual information for two divisions obtained by maximizing Z-modularity and modularity maximization. 

\begin{figure}[tb]
\centering
%% \subfloat[$n=1000$, $l=20$, and $p_\text{in}=0.5$.]{\includegraphics[width=0.49 \textwidth]{graph/ex1.pdf}}\\
%% \subfloat[$n=1000$, $l=50$, and $p_\text{in}=0.5$.]{\includegraphics[width=0.49 \textwidth]{graph/ex2.pdf}}\\
\includegraphics[width=0.49 \textwidth]{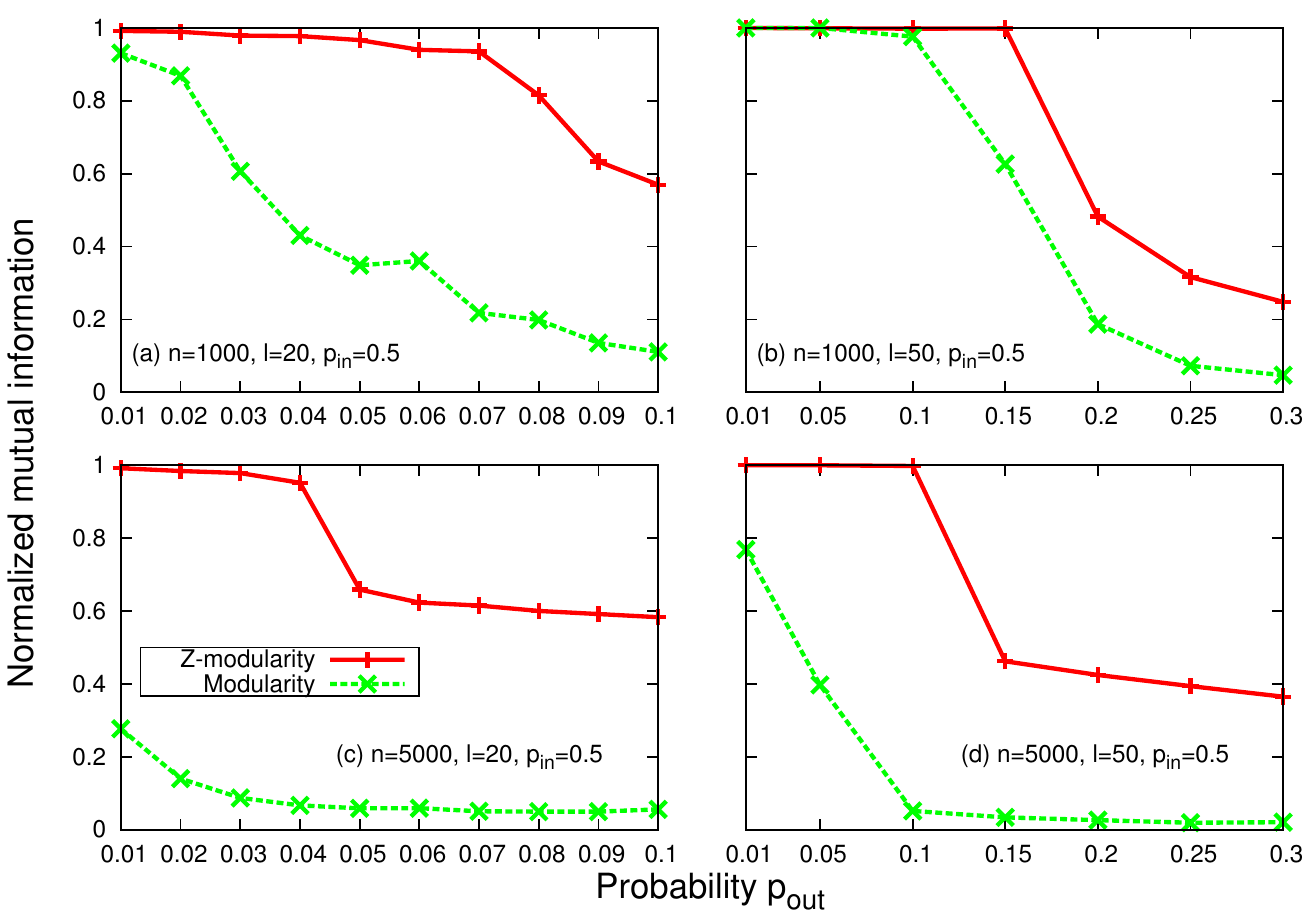}
\caption{(Color online) Results for the planted $l$-partition model.} \label{fig:ex2}
\end{figure}

The results are shown in Fig.~\ref{fig:ex2}. 
As can be seen, Z-modularity outperforms the original modularity in all four cases.
In particular, Z-modularity provides much more superior results compared to modularity for networks consisting of relatively small communities. 
%In particular, Z-modularity outperforms the modularity for networks consisting of  relatively small communities. 
%\TODO{add other parameter settings}

\medskip
\paragraph*{LFR benchmark.}
In the planted $l$-partition model, each group in a generated network forms the Erd\H{o}s--R\'{e}nyi random graph~\cite{ErRe59}. 
Thus, all vertices have approximately the same degree. 
Moreover, all groups have exactly the same size. 
These phenomena are rarely observed in networks in real-world systems. 
As a more realistic model, the LFR benchmark was proposed by Lancichinetti, Fortunato, and Radicchi~\cite{LaFoRa08} 
for the case of unweighted and undirected networks. 
The LFR benchmark was then extended to the case of directed and weighted networks with overlapping communities~\cite{LaFo09_1}. 
We now use the original unweighted and undirected case without overlapping communities. 

In the model, degree distribution and community size distribution follow the power law 
with exponents $\gamma$ and $\beta$, respectively. 
Furthermore, we can specify the number of vertices $n$, average degree $\langle k\rangle$, maximum degree $k_{\max}$, 
minimum community size $c_\text{min}$, maximum community size $c_\text{max}$, and mixing parameter $\mu$. 
In particular, mixing parameter $\mu$ indicates the mixing ratio of communities, 
i.e., the higher $\mu$ is, the more densely connected the communities are. 
The model constructs a network consistent with the specified parameters. 
For more details, see Ref.~\cite{LaFoRa08}. 
In our experiments, we set the parameters the same as used in Refs.~\cite{LaFo09,LaFo14} as follows: 
$\gamma=-2$, $\beta=-1$, $\langle k\rangle=20$, and $k_{\max}=50$. 
We construct four networks corresponding to combinations of two different network sizes ($n=1000$ or 5000) 
and two different ranges of community size ($(c_\text{min}, c_\text{max}) = (10,50)$ or (20,100)).
%($c_\text{min}=10$ and $c_\text{max} = 50$, or $c_\text{min}=20$ and $c_\text{max}=100$).

%the number of vertices $n$, power-law exponent $\tau_1$, power-law exponent $\tau_2$, 

\begin{figure}[tb]
\centering
\includegraphics[width=0.49 \textwidth]{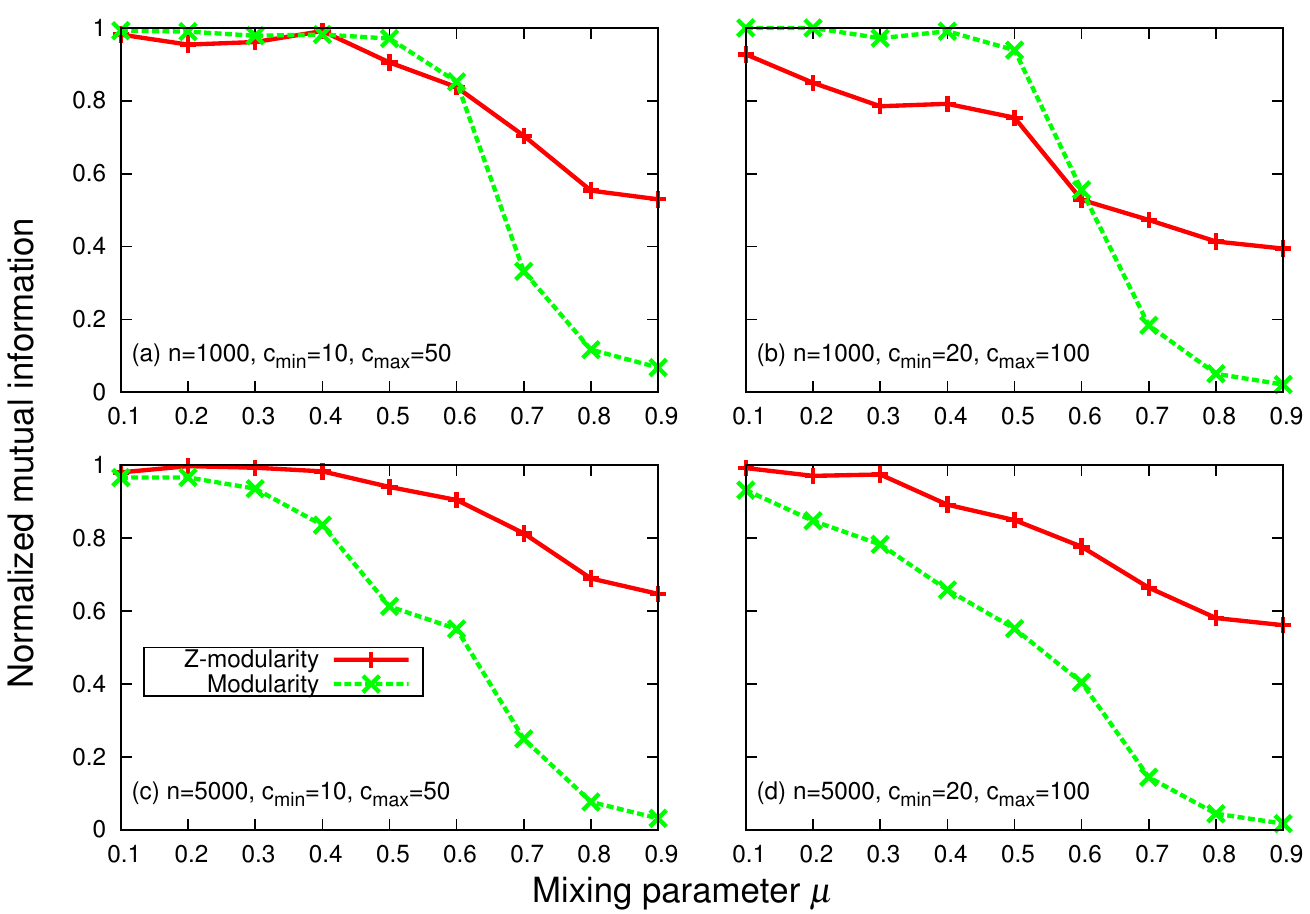}
\caption{(Color online) Results for the LFR benchmark.} \label{fig:ex3}
\end{figure}

The results are illustrated in Fig.~\ref{fig:ex3}. 
%For the smaller networks ($n=1000$), the mutual information values obtained by maximizing Z-modularity are lower than those obtained by modularity maximization when $\mu \leq 0.6$, whereas are higher when $\mu \geq 0.7$, in both community size settings. 
For the smaller networks ($n=1000$), the mutual information values obtained by maximizing Z-modularity are lower 
than those obtained by modularity maximization when $\mu \leq 0.6$ for both community size settings. 
This trend is significant when the network consists of relatively large communities ($(c_\text{min}, c_\text{max}) = (20,100)$).
On the other hand, for larger networks ($n=5000$), Z-modularity outperforms the original modularity for both community size settings. 
From the above, we see that Z-modularity is particularly suitable for identifying community structure 
when a network consists of relatively small communities. 

\begin{figure}[tb]
\centering
\subfloat[Ground-truth]{\includegraphics[width=0.23 \textwidth]{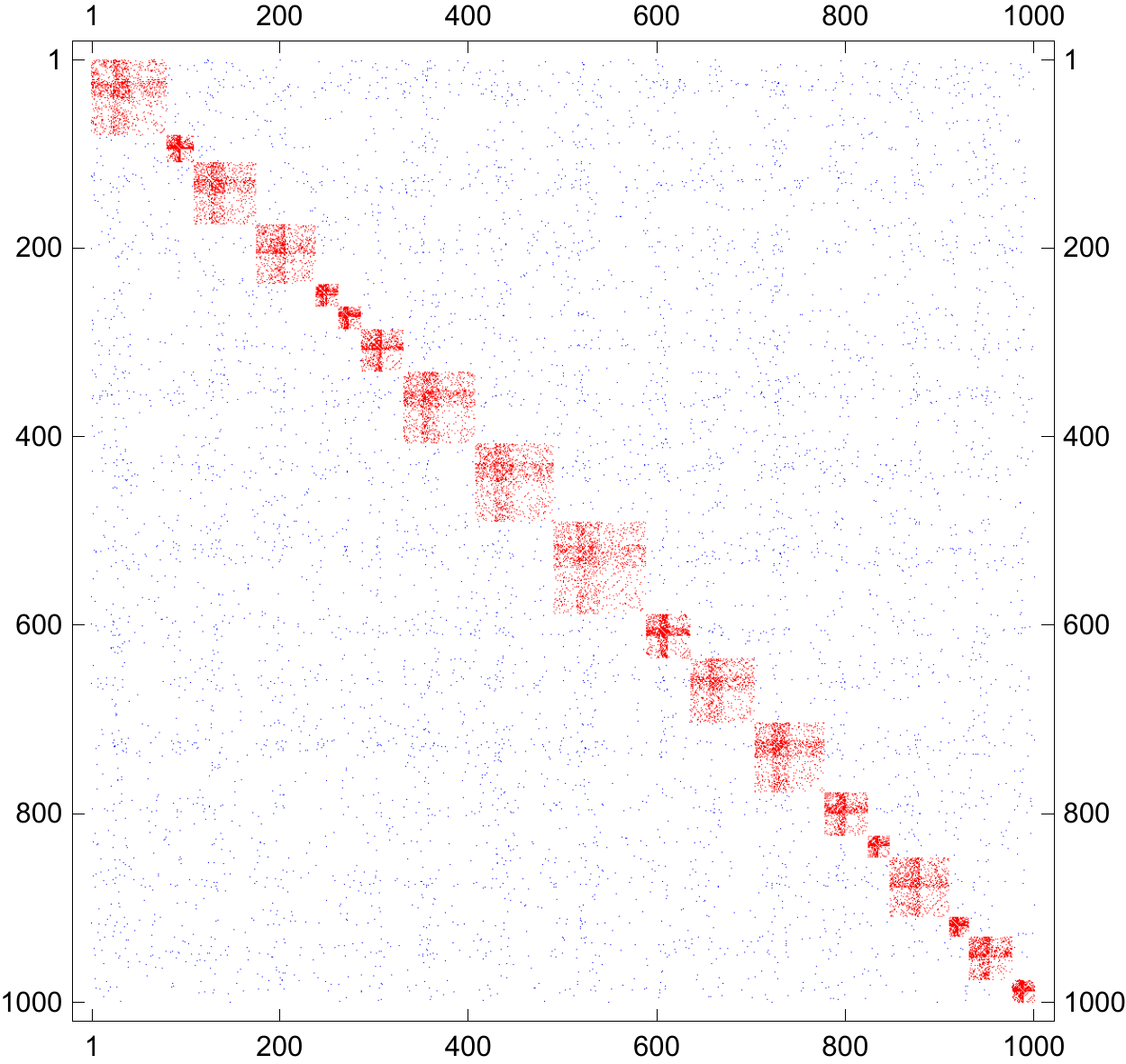}}\quad
\subfloat[Z-modularity]{\includegraphics[width=0.23 \textwidth]{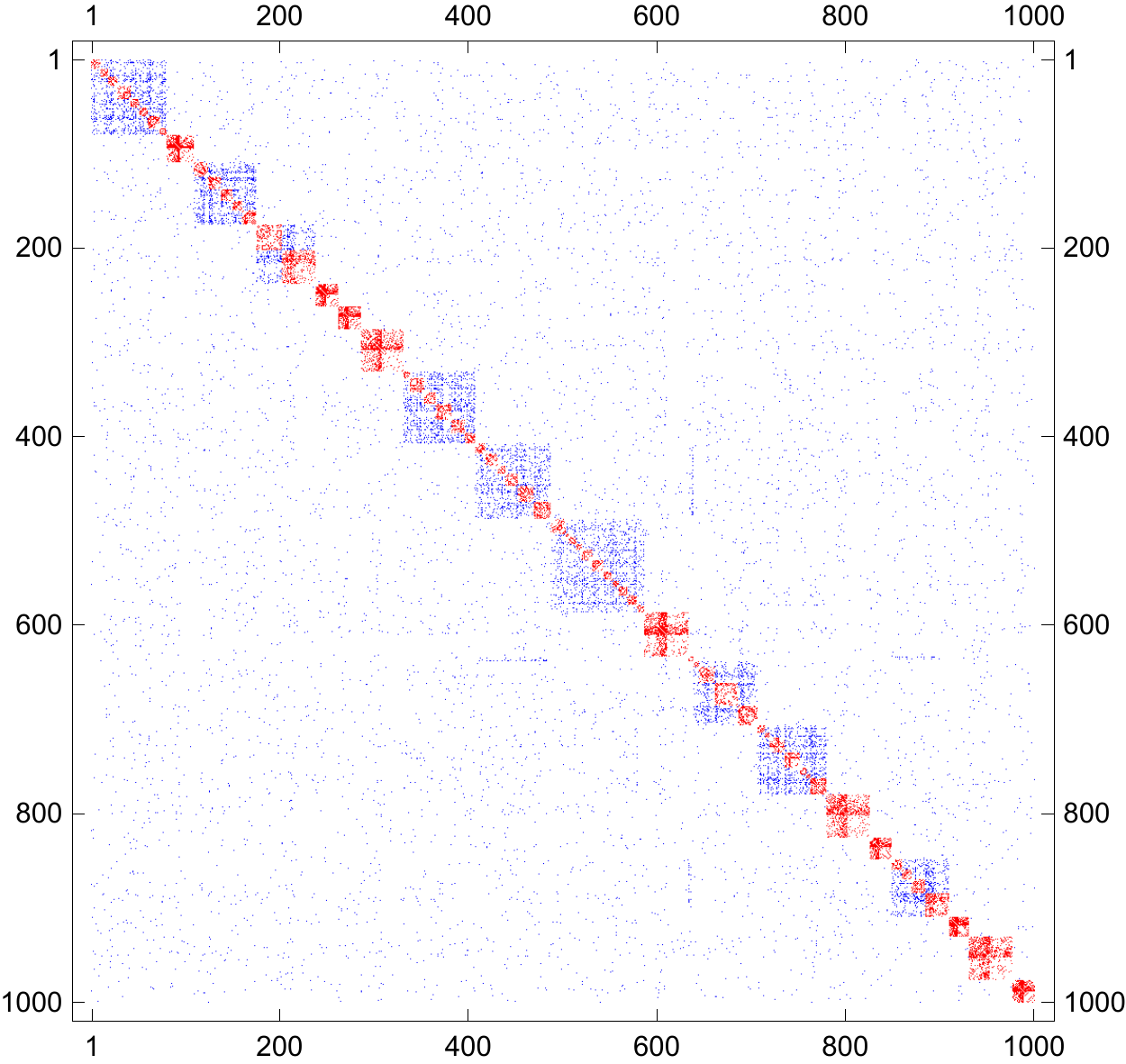}}
\caption{(Color online) Adjacency matrices for an LFR benchmark network.} \label{fig:adj}
\end{figure}

Here we investigate why the mutual information values obtained by maximizing Z-modularity are low 
when the community sizes are large. 
To this end, Fig.~\ref{fig:adj} depicts the adjacency matrices of the LFR benchmark network 
with parameters $\gamma=-2$, $\beta=-1$, $n=1000$, $\langle k\rangle=20$, $k_{\max}=50$, $c_{\min}=20$, $c_{\max}=100$, and $\mu=0.3$.
The vertices are ordered according to both the ground-truth partition and the optimal partition for Z-modularity.
The edges connecting vertices in the same community and in different communities are plotted with different colors, 
i.e., red and blue, respectively.
As can be seen, maximizing Z-modularity divides the relatively large ground-truth communities 
because they contain much denser communities in the hierarchical structure by random behavior. 
%there exists hierarchical structure by random behavior.

%% \begin{figure}[tb]
%% \centering
%% \subfloat[$n=1000$, $\langle k\rangle=20$, $k_{\max}=50$, $c_{\min}=10$, and $c_{\max}=50$.]{\includegraphics[width=0.23 \textwidth]{graph/lfr7.pdf}}\quad
%% \subfloat[$n=1000$, $\langle k\rangle=20$, $k_{\max}=50$, $c_{\min}=20$, and $c_{\max}=100$.]{\includegraphics[width=0.23 \textwidth]{graph/lfr6.pdf}}\\
%% \subfloat[$n=5000$, $\langle k\rangle=20$, $k_{\max}=50$, $c_{\min}=10$, and $c_{\max}=50$.]{\includegraphics[width=0.23 \textwidth]{graph/lfr5.pdf}}\quad
%% \subfloat[$n=5000$, $\langle k\rangle=20$, $k_{\max}=50$, $c_{\min}=20$, and $c_{\max}=100$.]{\includegraphics[width=0.23 \textwidth]{graph/lfr8.pdf}}
%% \caption{(Color online) Results for the LFR benchmarks.} \label{fig:ex3}
%% \end{figure}

\medskip
\paragraph*{Hanoi graph.}
% http://www.tp.umu.se/~rosvall/code.html
Here we demonstrate optimal partitions with respect to Z-modularity and the original modularity for the Hanoi graph, 
which is an example of networks with hierarchical organization.
The Hanoi graph $H_n$ corresponds to the allowed moves in the \textit{tower of Hanoi} for $n$ disks, 
which is a famous puzzle invented by {\'E}douard Lucas in 1883.
The Hanoi graph $H_n$ has $3^n$ vertices and $3\cdot (3^n-1)/2$ edges.
In the context of community detection in networks, the Hanoi graph $H_3$ is used by Rosvall and Bergstrom~\cite{RoBe11}.

\begin{figure}[tb]
\centering
\subfloat[Optimal partition for Z-modularity: 27 communities, $Z=3.376$, and $Q=0.6379$.]{\includegraphics[width=0.47 \textwidth]{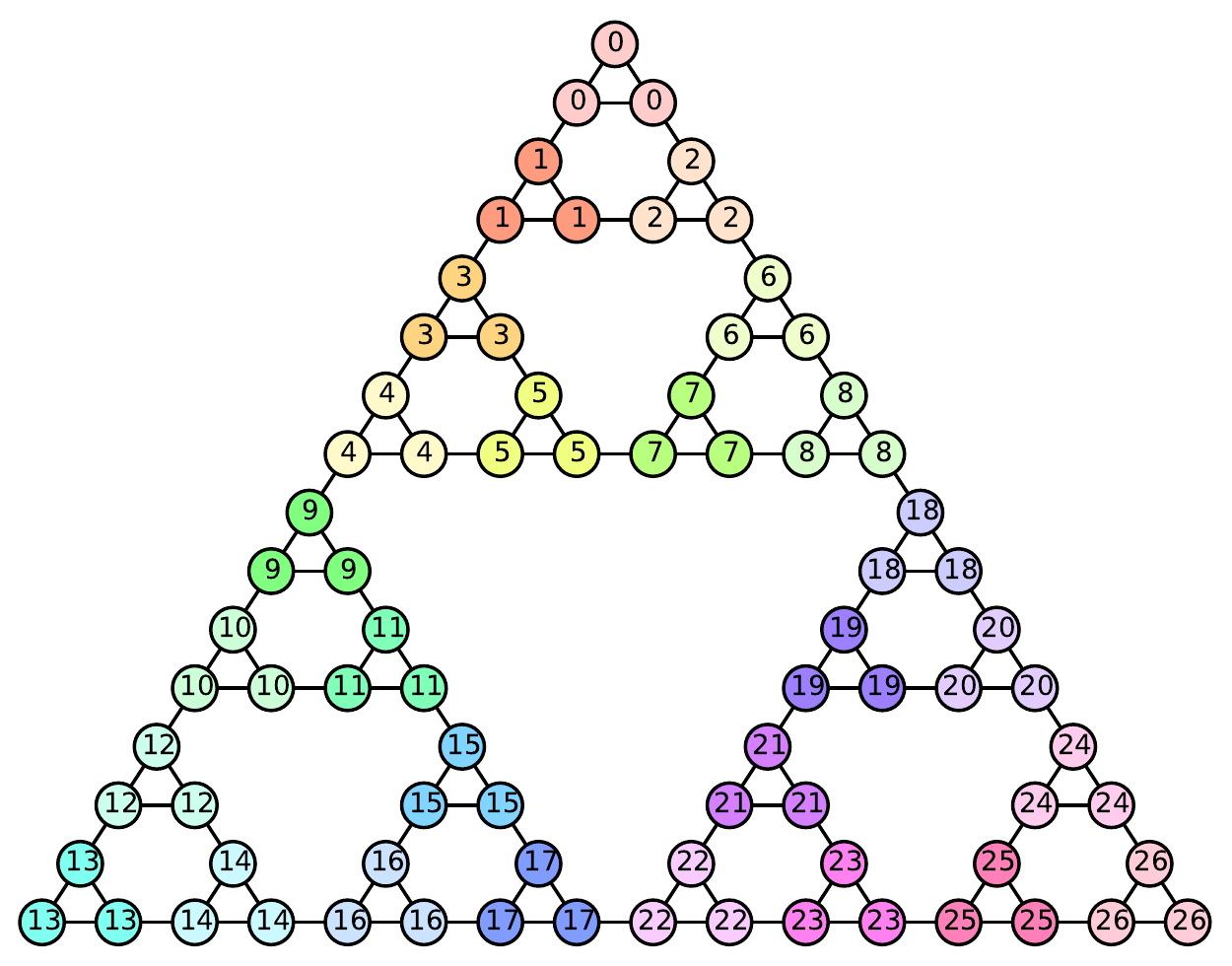}}\\
\subfloat[Optimal partition for modularity: 9 communities, $Z=2.510$, and $Q=0.7889$.]{\includegraphics[width=0.47 \textwidth]{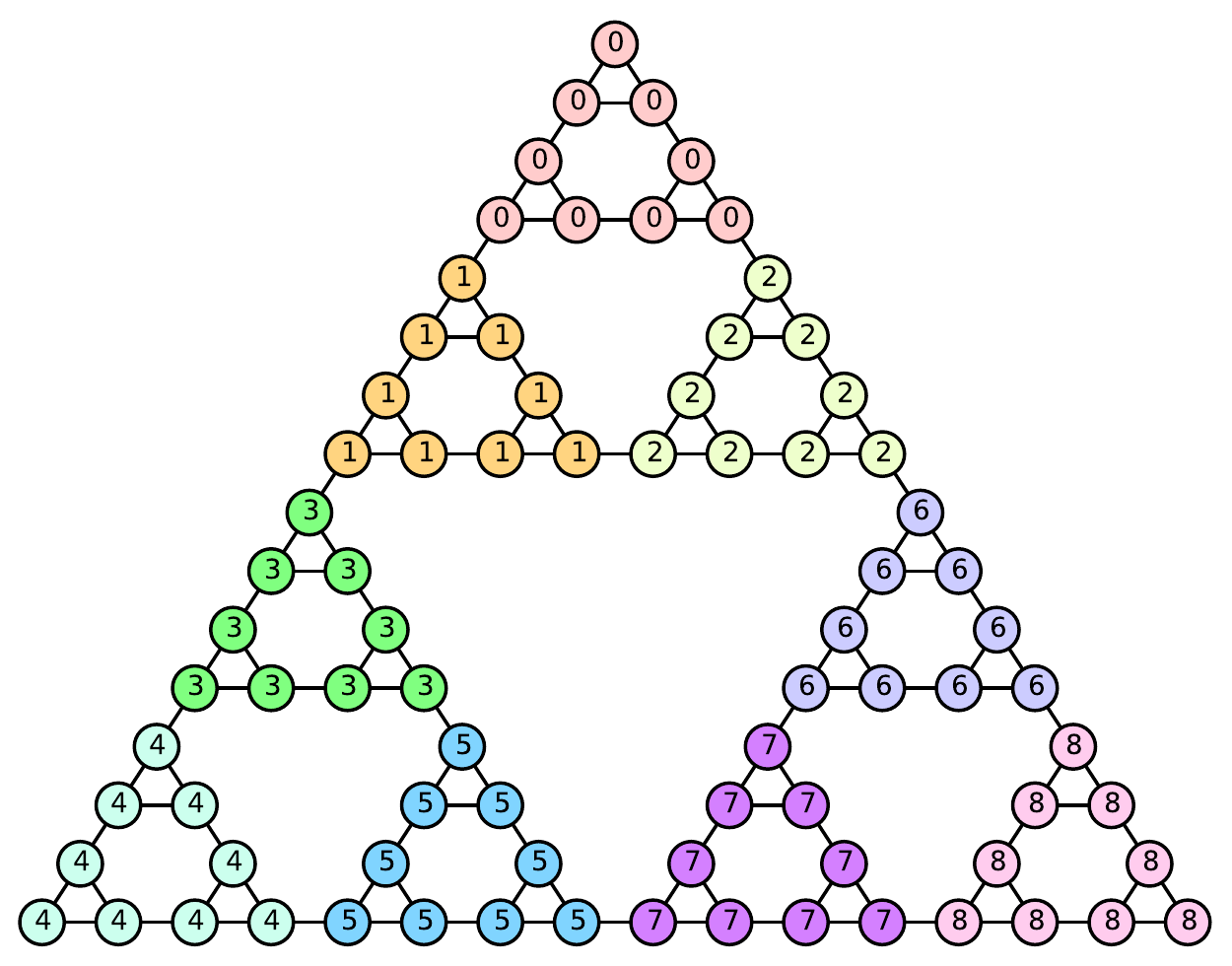}}\\
\caption{(Color online) Community structure for Hanoi graph $H_4$.} \label{fig:hanoi}
\end{figure}

The results for Hanoi graph $H_4$ are shown in Fig.~\ref{fig:hanoi}, 
where the label (and color) of each vertex represents the community to which the vertex belongs. 
As can be seen, maximizing Z-modularity leads to more detailed partition than modularity maximization. 
%In fact, modularity maximization divides the network into 9 $H_2$s, whereas maximizing Z-modularity divides the network into 27 $H_1$s. 

\subsection{Real-world networks}
Here we report the results of computational experiments with real-world networks; 
i.e., the Zachary's karate club network, the Les Mis\'{e}rables network, and the American college football network. 

\medskip
\paragraph*{Zachary's karate club network.}
The first example is the famous karate club network analyzed by Zachary~\cite{Za77}, 
which is often used as a benchmark to evaluate community detection methods. 
It consists of 34 vertices representing the members in a karate club in an American university, 
in addition to 78 edges representing friendship relations among individuals. 
Because of a conflict between the club administrator and the instructor, 
the club members split into two groups, one supporting the administrator and the other supporting the instructor. 
Therefore, these groups can be viewed as a ground-truth community structure. 

\begin{figure}[tb]
\centering
\includegraphics[width=0.49 \textwidth]{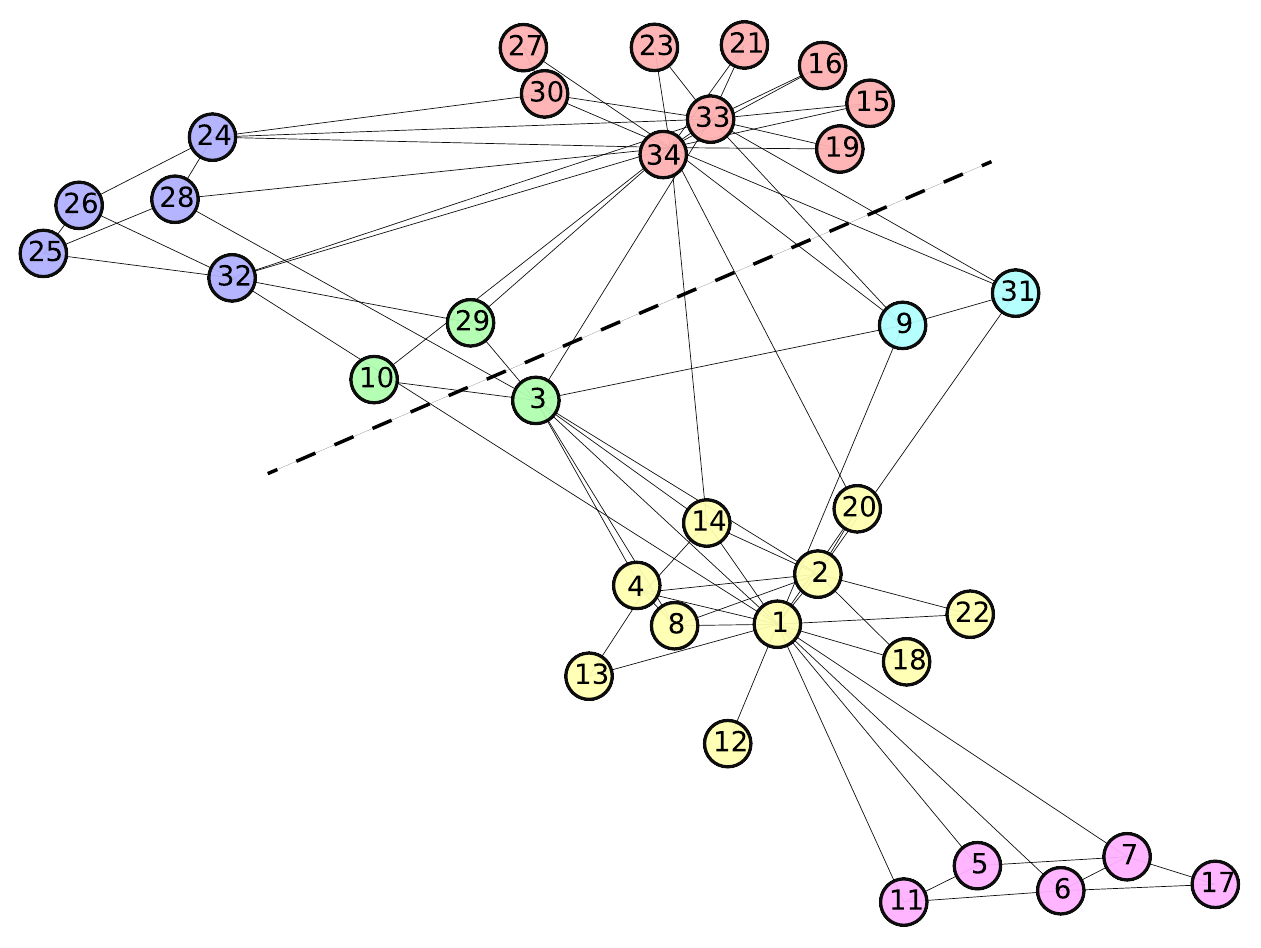}
\caption{(Color online) Community structure for Zachary's karate club network: 6 communities, $Z=0.9266$, and $Q=0.3882$.} \label{fig:karate}
\end{figure}

The division obtained by maximizing Z-modularity is shown in Fig.~\ref{fig:karate}, where vertices with the same color represent a community. 
The label of each vertex represents an identification number of the member. 
For example, 1 and 34 represent the administrator and the instructor, respectively. 
The dashed line gives the division of the network into the above two groups. 
Although the community $\{3,10,29\}$ straddles two groups, the other communities are all contained in either one of the groups. 

%% \medskip
%% \paragraph*{Lusseau's dolphins network.}
%% The second example is the network of bottlenose dolphins living in Doubtful Sound, New Zealand, analyzed by Lusseau~\cite{Luetal03}. There are 62 vertices representing the dolphins, and 159 edges connecting dolphins that were seen together more frequently expected by chance. 

%% The division obtained by maximizing Z-modularity is illustrated in Fig.~\ref{fig:dolphins}
%% \TODO{explain result}

%% \begin{figure}[tb]
%% \centering
%% \includegraphics[width=0.48 \textwidth]{img/dolphins.pdf}
%% \caption{(Color online) Community structure for Lusseau's dolphins network: 13 communities, $Z=1.42$, $Q=0.428$.} \label{fig:dolphins}
%% \end{figure}

\medskip
\paragraph*{Les Mis\'{e}rables network.}
The second example is the network of the characters in the novel \textit{Les Mis\'{e}rables} by Victor Hugo, compiled by Knuth~\cite{Kn93}. It consists of 77 vertices representing the characters and 254 edges indicating the co-appearance of characters. 

\begin{figure}[tb]
\centering
\includegraphics[width=0.49 \textwidth]{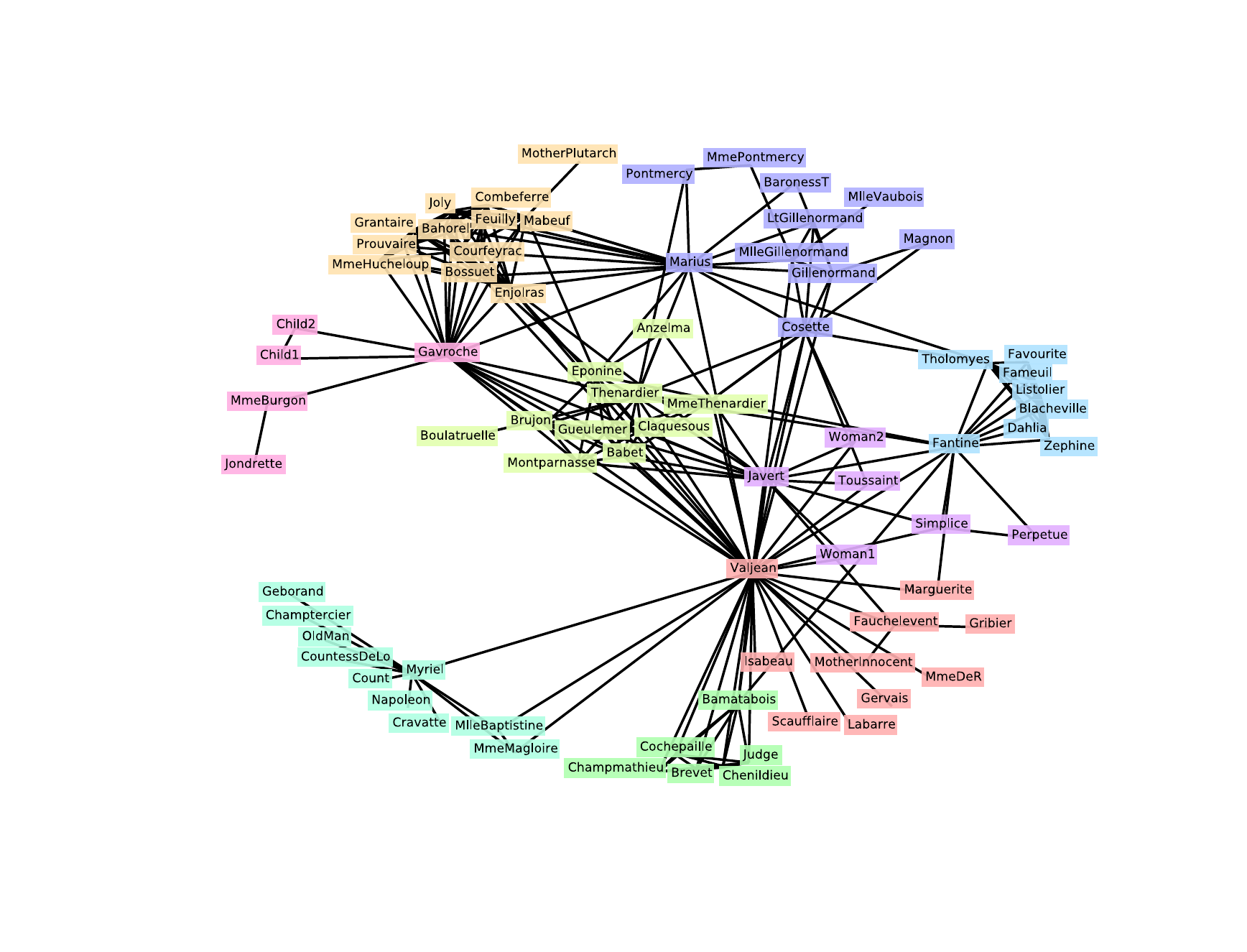}
\caption{(Color online) Community structure for Les Mis\'{e}rables network: 9 communities, $Z=1.490$, and $Q=0.5245$.} \label{fig:miserable}
\end{figure}

The division obtained by maximizing Z-modularity is presented in Fig.~\ref{fig:miserable}, 
where vertices with the same color represent a community. 
The label of each vertex represents the name of the character. 
Identified communities are likely to correspond to specific groups within the story. 
For example, the community consisting of 12 vertices (shaded with light brown) at the top left corner contains major characters belonging to the revolutionary student club \textit{Friends of the ABC}. 

\medskip
\paragraph*{American college football network.}
The third and final example is a network of college football teams in the United States, 
which was derived by Girvan and Newman~\cite{GiNe02}. 
There are 115 vertices representing the football teams, and 654 edges connecting teams that played each other in a regular season. The teams are divided into 12 groups referred to as \textit{conferences} containing approximately 10 teams each. 
More games are played between teams in the same conference than between teams in different conferences. 
Thus, the conferences can be viewed as a ground-truth community structure. 

\begin{figure}[tb]
\centering
\includegraphics[width=0.49 \textwidth]{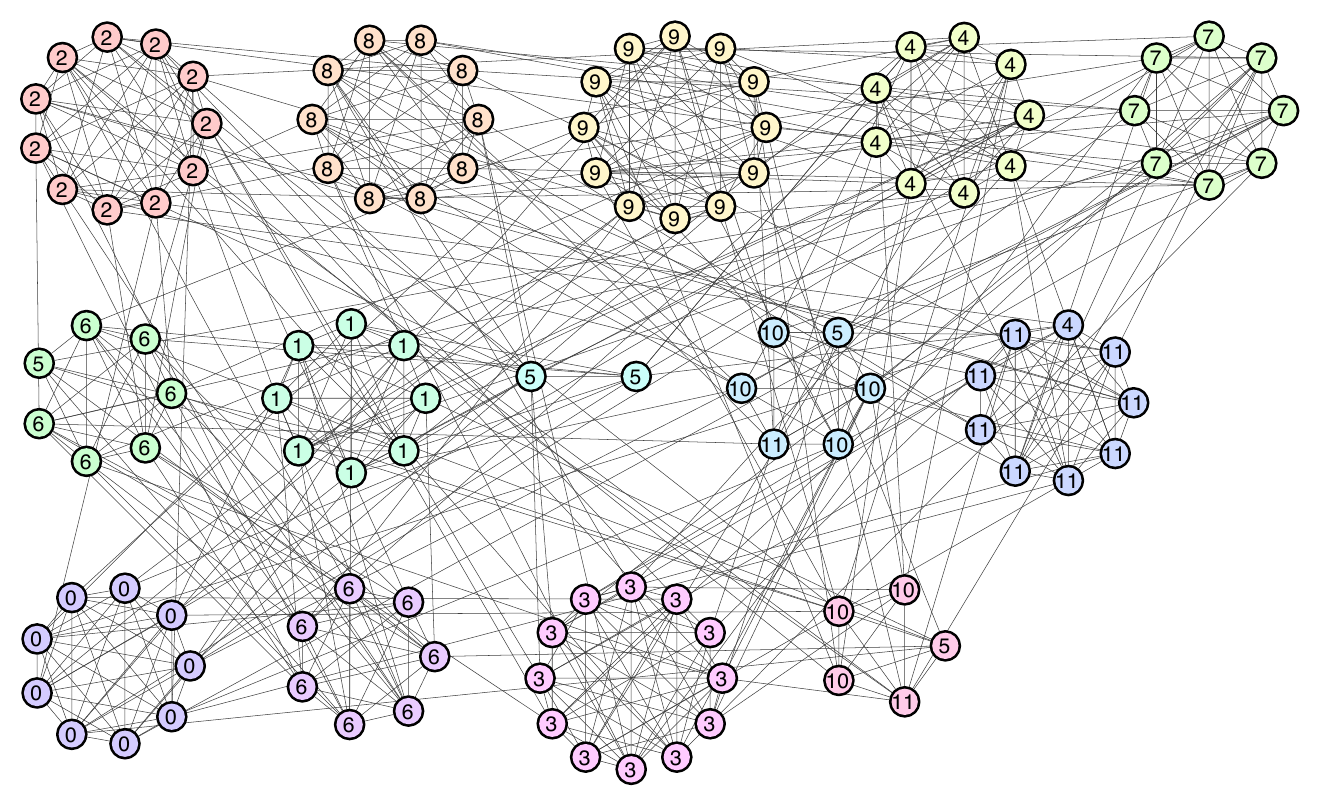}
\caption{(Color online) Community structure for American college football network: 14 communities, $Z=2.111$, and $Q=0.5738$.} \label{fig:football}
\end{figure}

The division obtained by maximizing Z-modularity is shown in Fig.~\ref{fig:football}, 
where vertices with the same color represent a community. 
Note that the label of each vertex now represents the conference to which the team belongs 
rather than an identification number of the team. 
Although some misclassifications are observed, 
Z-modularity correctly identifies 7 out of 12 conferences (i.e., conferences 0, 1, 2, 3, 7, 8, and 9). 
This result is outstanding in comparison with divisions obtained by modularity maximization.  
In fact, as reported in Ref.~\cite{AgKe08}, 
only four conferences were correctly recovered by division with a higher modularity value $Q=0.6046$.

\section{Conclusions}\label{sec:conclusion}

%In this study, we have discussed the normalization of the modularity using the standard deviation. 
%We first pointed out that the modularity overlooks the variance of the probability distribution of the fraction of the number of edges within communities under the division fixed. 
%This simple but critical observation brought our quality function Z-modularity. It measures Z-score of the division with respect to the fraction of the number of edges within communities. 
In this study, we have identified a problem in the concept of modularity and suggested a solution to overcome this problem. 
Specifically, we have obtained a new quality function Z-modularity 
that measures the Z-score of a given division with respect to the fraction of the number of edges within communities. 
Theoretical analysis has shown that Z-modularity mitigates the resolution limit of the original modularity in certain cases. 
In fact, Z-modularity never merges  adjacent cliques in the well-known ring of cliques network with any number and size of cliques. %%%%%%%added
%Indeed, Z-modularity correctly identifies natural community structure of the ring of cliques network with any number and size of cliques. 
%Indeed, it never merges adjacent cliques in the ring of cliques network with any number and size of cliques. 
In computational experiments, we have evaluated the validity and reliability of Z-modularity. 
The results for artificial networks show that Z-modularity more accurately detects the ground-truth community structure 
than the original modularity in most cases. 
In particular, Z-modularity outperforms modularity for networks consisting of relatively small communities. 
Furthermore, the results for real-world networks demonstrate that 
Z-modularity leads to natural and reasonable community structure in practical use. 
Therefore, we conclude that Z-modularity could be another option for the quality function in community detection. 

In the future, further experiments should be conducted to examine the performance of Z-modularity in more details. 
Although strict experiments were conducted in the present study, other experimental settings are also possible. 
%As another future direction, it should be investigated whether some statistical mechanics hide behind maximizing Z-modularity. 
As another future direction, the physical interpretation of maximizing Z-modularity should be investigated. 
For example, it is known that modularity maximization can be interpreted as the problem 
of finding the ground state of a spin glass model~\cite{ReBo06}. 

%As another future direction, we should investigate the physical phenomena behind maximizing Z-modularity. 

\begin{acknowledgments}
The first author is supported by a Grant-in-Aid for JSPS Fellows (No.~26-11908). 
The second author is supported by a Grant-in-Aid for Research Activity Start-up (No.~26887014).
This work was partially supported by JST, ERATO, Kawarabayashi Large Graph Project. 
\end{acknowledgments}

\bibliography{zmod}
\bibliographystyle{apsrev4-1}

\end{document}